\documentclass[journal]{IEEEtran}

\ifCLASSINFOpdf
\usepackage[pdftex]{graphicx}
\usepackage{epstopdf}
\usepackage{graphicx}
\usepackage{color}
\usepackage{amsmath,amsfonts}

\usepackage{amsmath,bm}
\usepackage{accents}
\usepackage[noadjust]{cite}

\usepackage{tikz-timing}
\usepackage{tikz}
\usepackage{subfigure}

\newtheorem{theorem}{Theorem}
\newtheorem{definition}{Definition}

\newtheorem{proposition}{Proposition}
\newtheorem{corollary}{Corollary}
\newtheorem{remark}{Remark}
\newtheorem{example}{Example}

\newcommand{\diag}{{\rm \;diag}}
\newcommand{\sinc}{{\rm \;sinc}}
\newcommand{\mat}[1]{\mathbf{#1}}
\newcommand{\R}{\mathbb{R}}

\newcommand{\RS}{\mathbb{S}}

\usepackage{epstopdf}
\usepackage{graphics}

\usepackage{mathtools}
\DeclarePairedDelimiter{\ceil}{\lceil}{\rceil}
\DeclarePairedDelimiter\floor{\lfloor}{\rfloor}

\ifCLASSINFOpdf
\else
\fi

\begin{document}

\title{Partial Phase Cohesiveness in Networks of Kuramoto Oscillator Networks}

\author{Yuzhen~Qin, 
		Yu~Kawano,
		Oscar~Portoles,
		and Ming~Cao
        
\thanks{Y. Qin and M. Cao are with Engineering and Technology
	Institute (ENTEG), O. Portoles is with Department  of  Artificial  Intelligence  and  Cognitive  Engineering, University of Groningen, the Netherlands. \{emails: y.z.qin, o.portoles.marin, m.cao\}@rug.nl. Y. Kawano is with the Faculty of Engineering, Hiroshima University, Higashi-Hiroshima, Japan (email: ykawano@hiroshima-u.ac.jp). The work  was supported in part by the European Research Council (ERC-CoG-771687) and the Netherlands Organization for Scientific Research (NWO-vidi-14134). 
}	}

\maketitle

\begin{abstract}
	Partial, instead of  complete, synchronization has been widely observed in various networks including, in particular,  brain networks. Motivated by data from human brain functional networks,   in this technical note, we analytically  show that partial synchronization can be induced by strong regional connections in coupled subnetworks of Kuramoto oscillators. To quantify the  required strength of regional connections, we first obtain a critical value for the algebraic connectivity of the corresponding subnetwork using the incremental $2$-norm. 
	We then introduce the concept of the \textit{generalized complement graph}, and obtain another condition on the weighted nodal degree by using the incremental $\infty$-norm. Under these two conditions, regions of attraction  for partial phase cohesiveness are estimated in the forms of the incremental $2$- and $\infty$-norms, respectively.  Our result based on the incremental $\infty$-norm is the first known  criterion that is applicable to non-complete graphs. Numerical simulations are performed on a two-level network to illustrate our theoretical results; more importantly, we use real anatomical brain network data to show how our results may  reveal the interplay between anatomical structure and empirical patterns of synchrony. 
\end{abstract}

\section{Introduction}
 Neuronal synchronization across cortical regions of human brain, which has been widely detected through recording and analyzing brain waves, is believed to facilitate communication among neuronal ensembles \cite{womelsdorf2007modulation}, and only closely correlated oscillating neuronal ensembles can exchange information effectively \cite{fries2005mechanism}. 
In healthy human brain, it is frequently observed that only a part of its cortical regions are synchronized \cite{palva2012discovering},  and such a phenomenon is commonly referred to as partial phase cohesiveness or partial synchronization of brain neural networks.  In contrast, in pathological brain of  a patient such as  an epileptic,  complete synchronization of neural activities takes place across the entire brain \cite{fisher2005epileptic}.  These observations suggest that healthy brain has powerful regulation mechanisms that are not only able to render synchronization, but also capable of preventing unnecessary synchronization among neuronal ensembles.
Partly motivated by these experimental  studies, researchers are interested in theoretically studying cluster or partial synchronization \cite{pecora2014cluster,favaretto2017cluster,menara2018stability,tiberi2017synchronization} and chimera states \cite{cho2017stable}, even though analytical results are much more difficult to obtain, while analytical results for complete synchronization are ample, e.g.,  \cite{dorfler2011critical,dorfler2014synchronization,jadbabaie2004stability}.

In our research, our ultimate objective is to  identify a possible underlying mechanism of partial phase cohesiveness in human brain. Employing the Kuramoto model \cite{kuramoto1975self}, which has been widely used to describe the dynamics of coupled neural ensembles~\cite{schmidt2014dynamics,cabral2014exploring}, we analytically study how partial phase cohesiveness can occur in a network of coupled oscillators. In human brain, the organization of cortical neurons  exhibits a ``network-of-networks'' structure in the sense that a cortical region is typically composed of  strongly connected ensembles of neurons that interact not only locally but also with ensembles in other regions \cite{barreto2008synchronization}. As neural ensembles in a cortical region are adjacent in space, it is  thus reasonable to assume that oscillators  within a brain region are coupled  through an all-to-all network, forming local communities at the lower level; at the higher level, the communities are interconnected by a sparse network facilitated through bundles of neural fibers connecting regions of the brain.  Motivated by these facts, we consider in this note the networks describing the interaction between Kuramoto oscillators have this two-level structure. 

The main contributions of this note are some new sufficient conditions by using Lyapunov functions utilizing the incremental $2$-norm and $\infty$-norm, which ensure partial phase cohesiveness can take place in some subneworks of interest.
The incremental $2$-norm was first proposed in \cite{jadbabaie2004stability,chopra2009exponential}, in which some conditions for locally exponentially stable synchronization was obtained.  Later on, it was also employed in the study of non-complete  networks \cite{dorfler2012synchronization,dorfler2012exploring}.  Inspired by these works, we first employ the incremental $2$-norm and obtain a sufficient condition for the algebraic connectivity $\lambda_2(L)$ of the considered subnetwork, and then estimate the region of attraction and the ultimate boundedness of phase cohesiveness. This critical value for $\lambda_2(L)$ depends on the natural frequency heterogeneity of the oscillators within the subnetwork and the strength of the connections from  its outside to this subnetwork. Since the incremental $2$-norm depends greatly on the scale, the obtained critical value and the estimated region of attraction are both conservative, especially when there are large numbers of oscillators in the considered subnetwork.


On the other hand, the incremental $\infty$-norm is scale independent.  It is always utilized to prove the existence of phase-locking manifolds and their local stability.  Existing conditions are usually expressed implicitly by a combined measure \cite{dorfler2013synchronization,Jafarpour2018}, and the regions of attraction are not estimated \cite{menara2018stability,menara2019exact}. To the authors' best knowledge, the best result on explicit conditions utilizing the incremental $\infty$-norm is given in \cite{dorfler2011critical}, which has only studied unweighted complete networks.  It is challenging to extend it to the non-complete or even weighted complete networks. To meet the challenges, we introduce a concept of the \textit{generalized complement graph} in this note, which, in turn, enables us to make use of the incremental $\infty$-norm and obtain an explicit condition. Compared to the results obtained by the incremental $2$-norm: 1) the established sufficient condition is less conservative if the dissimilarity of natural frequencies and the strengths of external connections are noticeable; 2) more importantly,  the region of attraction we identified is much larger.
After simplifying the network structure, our results on partial phase cohesiveness can reduces to some results on complete phase cohesiveness. The reduced results are sharper than the best known result obtained by using incremental $2$-norm for the case of weighted complete and non-complete networks \cite[Theorem 4.6]{dorfler2012exploring} (especially in terms of the region of attraction), and are identical to the sharpest one found in \cite{dorfler2011critical} for the case of unweighted complete networks. The only drawback of our condition is that each oscillator is required to be connected to a minimum number of other oscillators.
Finally, we perform some simulations using the anatomical brain network data obtained in \cite{finger2016modeling}; the simulation results show how our theoretical findings may reveal a possible mechanism that gives rise to various patterns of synchrony detected in empirical data of human brain \cite{portoles2018characterizing}. Our preliminary work was presented in \cite{qin2018ECC}, where only the incremental $2$-norm was studied. Moreover, we consider a more general inter-community coupling structure in this note, without requiring  that every node in one community is connected to all the nodes in another. 


%

The rest of this note is organized as follows. We introduce the model on the two-level networks and formulate the  problem of partial phase cohesiveness in Section \ref{Sec_PF}. The first result is obtained by using the incremental $2$-norm in Section \ref{2-norm}. Section \ref{inft-norm} introduces the notion of generalized complement graphs and derives the main result utilizing the incremental $\infty$-norm. Some simulations are performed in Section \ref{simulation}. 

\textit{Notations}:  Let $\mathbb R$ and $\mathbb R_{\ge 0}$ be the set of real numbers and nonnegative real numbers, respectively.  For any positive integer $n$, let $\mathcal{T}_n:=\{1,2,\dots,n\}$, and $\mathbf{1}_n$ be the all-one vector. Denote the unit circle by $\RS^1$, and a point on it is called a \textit{phase} since the point can be used to indicate the phase angle of an oscillator.  {For any two phases $\gamma_1, \gamma_2 \in \mathbb S^1$, the geodesic distance between them is the minimum of the lengths of the counter-clockwise and clockwise arcs connecting them, which is denoted by $|\gamma_1-\gamma_2|$. Note that $|\gamma_1-\gamma_2|\le \pi$ for any $\gamma_1, \gamma_2 $.} Let $\mathbb{T}^n:=\RS^1\times\cdots \times \RS^1$ denote the $n$-torus.  For any $x \in \R$, Let $\floor x$ denote the largest integer that is less than or equal to $x$, and $\ceil x$ the smallest integer that is greater than or equal to $x$. Let $\|\cdot\|_p$  denote the $p$-norm for both vectors and matrices, where $p\ge 1$ can be infinite.


\section{Problem Setup}\label{Sec_PF}

We consider a network of $M>1$ communities, each of which consists of $N\ge 1$ fully connected heterogeneous Kuramoto oscillators.  The graph of the network, which describes which community is interconnected to which other communities, is in general not a complete graph. The dynamics of the oscillators are described by 
\begin{align}
\dot \theta_i^p&=\omega_i^p+{K^p} \sum\nolimits_{ n=1}^{N} \sin(\theta^p_{n}-\theta_i^p) ,\nonumber \\
+&\sum\nolimits_{q=1}^{M}  \sum\nolimits _{n=1}^{N} {a^{p,q}_{i, n}}\sin(\theta^q_{ n}-\theta_i^p),{p \in \mathcal T_M, i \in \mathcal T_N,} \label{model}
\end{align}
where $\theta_i^ p \in \RS^ 1$ and $\omega_i^ p \ge 0$ represent the phase and natural frequency of the $i$th oscillator in the $p$th community, respectively. Here, the uniform coupling strength of all the edges in the complete graph of the $p$th community is denoted by $K^p>0$, which we refer to as  the \textit{local} coupling strengths. 
The coupling strengths $a^{p,q}_{i, n}$,  which we call the \textit{inter-community} coupling strengths,  satisfy  $ a^{p,q}_{i, n}>0$ if $i\neq n$ and there is a connection between the $i$th oscillator in the $p$th community and the $j$th oscillator in the $q$th community, and $a^{p,q}_{i,n}=0$ otherwise.  We define the inter-community coupling matrices by $A^{p,q}: = [a^{p,q}_{i,j}] _{N\times N}\in \mathbb R^{N\times N}$, and each satisfies $A^{p,q}=A^{q,p}$. 

\begin{remark}
	 Our analysis later on applies to the case when each community has a different network topology and even when the numbers of oscillators in the communities are different. However, for the sake of notational simplicity,  we assume that each community is connected by a uniformly weighted complete network.
\end{remark}

 {The Kuramoto oscillator network model~\eqref{model} is used in~\cite{schmidt2014dynamics} to study synchronization phenomena of human brain. Along this line of research and motivated by brain research data, we focus on studying the widely observed but still not well understood phenomenon  for networks of communities of Kuramoto oscillators, the so called \textit{partial phase cohesiveness}, in which some but not all of the oscillators have close phases. 
 To facilitate the discussion of some  properties of interest for a subset of communities in the network, we  use $\mathcal T_r=\{1,\dots,r\}$, $1\le r \le M$,  to denote a set of chosen communities with the aim to investigate how phase cohesiveness can occur among these $r$ communities. We then define the following set to capture the situation when the oscillators in the communities in $\mathcal T_r$ reach phase cohesiveness. 
\begin{definition}\label{defi_cohesiveness}
	 Let $\theta \in \mathbb T^{MN}$ be a vector composed of the phases of all $N$ oscillators in all $M$ communities. Then, for  a given $\mathcal T_r$ and $\varphi\in [0,\pi]$, define the \textit{partial phase cohesiveness set}:
	\begin{align}\label{S_phi}
	\mathcal S_\infty(\varphi):=\left\{\theta\in\mathbb{T}^{MN}: \max_{i,j \in \mathcal T_N,k,l\in \mathcal T_r}|\theta_i^k-\theta_j^l|\le \varphi \right \}.
	\end{align}	
\end{definition}

Note that  $\varphi$ describes a level of phase cohesiveness since it is the maximum pair-wise phase difference of the oscillators in $\mathcal T_r$.  The smaller $\varphi$ is, the more cohesive the phases are.  All the phases in $\mathcal T_r$ are identical when $\varphi=0$, which is called \textit{partial phase synchronization}, and this can only happen when all the oscillators have the same natural {frequency}. In this paper, we allow the natural frequencies to be different, and are only interested in the cases when phase differences in $ \mathcal T_r$ are small enough.  We say that partial phase cohesiveness can take place  in $\mathcal T_r$ if  the solution ${\theta}:\R_{\ge 0}\to \mathbb{T}^{MN}$ to the system \eqref{model}   asymptotically converges to this set $\mathcal S_\infty({\varphi})$ for some $\varphi\in[0,\pi/2)$. We call the particular case when $\mathcal T_r=\mathcal T_M$  \textit{ complete} phase cohesiveness, which is also called practical phase synchronization  in \cite{dorfler2014synchronization}. In the rest of the paper, we study the partial phase cohesiveness by investigating how a solution $\theta(t)$ can  asymptotically converge to the set $\mathcal S_\infty({\varphi})$ and also by estimating the value of $\varphi$.

Let  $\mathcal{G}_{r} = (\mathcal{V}_{r}, \mathcal{E}_{r}, Z)$ denote the subgraph composed of the nodes in the communities contained in $\mathcal T_r$ and the edges connecting pairs of them. The weighted  adjacency matrix of this subgraph $Z:=[z_{ij}]_{Nr \times Nr}\in \mathbb R^{Nr \times Nr}$ is then  given by
\begin{equation}\label{adjaceny_R}
Z := \left[ {\begin{array}{*{20}{c}}
	{{K^1}C}&{A^{1,2}}& \cdots &{A^{1,r}}\\
	{A^{1,2}}&{{K^2}C}& \cdots &{A^{2,r}}\\
	\vdots & \vdots & \ddots & \vdots \\
	{A^{1,r}}&{A^{2,r}}& \cdots &{{ K^r}C}
	\end{array}} \right],
\end{equation}
where $C=[c_{ij}]_{N \times N} \in {\mathbb R}^{N \times N}$ is the adjacency matrix of a complete graph with $N$ nodes, where $c_{ij}=1$ for $i\neq j$, and $c_{ij}=0$ otherwise (recall that $A^{p,q}$ is symmetric). Let $D:=\diag(Z \mathbf 1_{Nr})$, then the Laplacian matrix of the graph $\mathcal G_r$ is $L:=D-Z $. Let $\lambda_2(L)$ {denote} the second smallest eigenvalue of $L$, which is always referred to as the \textit{algebraic connectivity} \cite{godsil2013algebraic}.

Let $\theta^p:=[\theta^p_1,\dots,\theta^p_N]^\top$, $\omega^p:=$ $[\omega^p_1,\dots,\omega^p_N]^\top$ for all $p \in \mathcal T_M$. As we are only interested in the behavior of the oscillator in $\mathcal G_r$, we define $x: = [{\theta^1}^\top,  \dots,{\theta^r}^\top]^\top $, and $\varpi  := [{\omega^1}^\top,\dots,{\omega^r}^\top]^\top$. For $i\in \mathbb{N}$, we define $\mu(i):=\ceil{i/N}$ and $\rho(i):=i-N \cdot \floor{i/N}$. By using these new notations, from (1), the dynamics of the oscillators on $\mathcal G_{r}$  can be rewritten as 
\begin{align}\label{dyna:xi}
{{\dot x }_i}=\varpi_i +&\sum\nolimits_{n = 1}^{Nr} {z_{i,n}}\sin ( {x _n} - {x _i} ) \nonumber\\
&+\sum\nolimits_{q=r+1}^{M} \sum\nolimits_{n=1}^{N} { a^{\mu(i),q}_{ \rho(i),n}}\sin(\theta_n^q-x_i),
\end{align}
where $i\in {\mathcal T_{Nr}}$. The first summation term describes the interactions among the oscillators within the subset of communities $ \mathcal T_r$, and the second one represents the interactions {from the outside of $ \mathcal T_r$ to the oscillators in $ \mathcal T_r$.}
In order to study the phase cohesiveness of the oscillators in $ \mathcal G_r$, we then look into the dynamics of pairwise phase differences, given by
\begin{align}
{{\dot x }_i} &- {{\dot x }_j} = {\varpi _i} - {\varpi_j}   \nonumber\\
&+ \sum\nolimits_{n = 1}^{Nr} \left({{z_{i,n}}\sin \left( {{x _n} - {x _i}} \right)} - {{z_{jn}}\sin \left( {{x _n} - {x _j}} \right)} \right) + u_{ij},\nonumber \\
&u_{ij}:=\sum\nolimits_{q=r+1}^{M}  \sum\nolimits_{n=1}^{N} \Big(a^{\mu(i),q}_{\rho(i),n}\sin(\theta_n^q-x_i) \nonumber \\ 
&\;\;\;\;\;\;\;\;\;\;\;\;\;\;\;\;\;\;\;\;\;\;\;\;\;\;\;\;\;\;\;\;\;\;\;\;\;\;- {a^{\mu(j),q}_{\rho(j),n}} \sin(\theta_n^q- x_j)\Big),\label{incremental}
\end{align}
where $i,j\in {\mathcal T_{Nr}}$. Let $\mat{u}_r:=[u_{ij}]_{i<j}\in \mathbb R^ {Nr(Nr-1)/2}$.
The incremental dynamics \eqref{incremental} play crucial roles in what follows. In the next two sections, we study partial phase cohesiveness in $\mathcal G_r$ with the help of \eqref{incremental}  using the incremental $2$-norm or $\infty$-norm (which will be introduced subsequently). To analyze phase cohesiveness, the techniques of ultimate boundedness theorem \cite[Theorem 4.18]{khalil2002nonlinear} will be employed. 

\section{Incremental $2$-Norm}\label{2-norm}
In this section, we introduce the incremental $2$-norm, and use it as a metric to study partial phase cohesiveness. According to Definition \ref{defi_cohesiveness}, we observe that a partially phase cohesive solution across $ \mathcal T_r$ should satisfy $|x_i-x_j| \le \varphi$ for all $i,j\in { \mathcal T_{Nr}}$. A quadratic Lyapunov function has been widely used to study phase cohesiveness even when the graph is not complete \cite{chopra2009exponential, jadbabaie2004stability, dorfler2014synchronization,dorfler2012exploring}, which is defined by
\begin{align}\label{Theo:2-norm}
W(x) :=  \frac{1}{2}\|B_c^\top x\|_2^2,
\end{align}
where $B_c \in \R^{Nr \times (Nr(Nr-1)/2)}$ is the incidence matrix of the complete graph. It is also known as the incremental $2$-norm metric of phase cohesiveness. 
For a given $\gamma \in [0,\pi)$, define
\begin{align}\label{D(gamma)_def}
{ \mathcal S_2({\gamma})}:=\left \{\theta \in \mathbb{T}^{MN}: \|B_c^\top x\|_2 \le \gamma \right\}.
\end{align}
{Note that $\mathcal S_2({\gamma})\subseteq \mathcal S_\infty({\gamma})$ for any given $\gamma \in[0,\pi)$.}  Different from the existing results that apply to complete cohesiveness taking place among all the oscillators in the networks \cite{chopra2009exponential, jadbabaie2004stability, dorfler2014synchronization,dorfler2012exploring},  we have studied partial phase cohesiveness in our previous work \cite{qin2018ECC} using the incremental $2$-norm metric. Compared to our previous work, we consider more general inter-community coupling structures as stated in Section \ref{Sec_PF}. 

{Let $\hat B_c=|B_c|$ be the element-wise absolute value of the incidence matrix $B_c $.  Let  $d^{\rm ex}_i=\sum\nolimits_{m=r+1}^{M} \sum\nolimits_{n=1}^N { a^{\mu(i),m}_{\rho(i),n}}$ for all $i\in \mathcal T_{Nr}$, and denote $D^{\rm ex}:=[d^{\rm ex}_1,\dots,d^{\rm ex}_{Nr}]^\top$.}  Now let us provide our first  result on partial phase cohesiveness on incremental $2$-norm.   A similar result can be found in \cite[Theorem 4.4]{dorfler2012synchronization}. Difference from it, we consider a two-level network, i.e., communities of oscillators, and study the partial phase cohesiveness. 
\begin{theorem}\label{theo:2-norm}
	Assume that the algebraic connectivity of $\mathcal{G}_{r}$ is greater than the critical value  specified by
	\begin{equation}\label{cluster_condition}
 	\lambda_2(L)> \|B_c^\top \varpi \|_2+{\| \hat B_c^\top D^{\rm ex}\|_2}.
	\end{equation}
	Then, each of the following equations
		\begin{align*}
		&\lambda_2(L) \sin(\gamma_s)-{\| \hat B_c^\top D^{\rm ex}\|_2}=\|B_c^\top \varpi \|_2,\\
		&(\pi/2) \lambda_2(L)  \sinc(\gamma_m)-{\| \hat B_c^\top D^{\rm ex}\|_2} =\|B_c^\top\varpi\|_2,	
		\end{align*} 
		has a unique solution, $\gamma_s\in[0,\pi/2)$ and $\gamma_m\in(\pi/2,\pi]$,	
	respectively, where $\sinc (\eta) = \sin(\eta)/\eta$ for any $\eta \in  \mathbb S^1$. Furthermore, the following statements hold: 
	\begin{enumerate}
		\item[(i)] for any $\gamma\in[\gamma_s,\gamma_m]$, { $ \mathcal S_2({\gamma})$ is a positively invariant set of the system \eqref{model}};
		\item[(ii)] for any $\gamma\in [\gamma_s,\gamma_m)$, the solution to \eqref{model} starting from  any $\theta(0) \in \mathcal  S_2({\gamma})$  converges to the set $ \mathcal S_2({\gamma_s})$.
	\end{enumerate}
\end{theorem}
\begin{IEEEproof}
	Choose $W(x)$ in (\ref{Theo:2-norm}) as a Lyapunov candidate.  Similar to the proof of  \cite[Theorem 4.6]{dorfler2012exploring}, we take the time derivative of $W(x)$ along the solution to \eqref{model} and obtain
	\begin{align*}
	\dot W(x)&\le x^\top B_c B_c^\top \varpi\\
	&-\sinc (\gamma) Nr x^\top B_c \diag (\{z_{ij}\}_{i<j}) B_c^\top x 
	+ x^\top B_c  \mat{u}_r.
	\end{align*}	
	From \cite[Lemma 7]{dorfler2012synchronization}, it holds that $x^\top B_c   \diag (\{z_{ij}\}_{i<j})  B_c^\top x $ $\ge \lambda_2(L)\|B_c^\top x\|_2^2/(Nr)$. From the definition of $\mat{u}_r$, one can evaluate that $\|\mat{u}_r\|_2\le \| \hat B_c^\top D^{\rm ex}\|_2$.
	As a consequence, we arrive at
	\begin{align*}
	\dot W(x)\le  x^\top B_c B_c^\top \varpi - \lambda_2(L) &\sinc (\gamma) \|B_c^\top x\|_2^2\\
	&+ \|B_c^\top x\|_2 {\| \hat B_c^\top D^{\rm ex}\|_2}.
	\end{align*}
	Following similar steps as those in the proof of \cite[Theorem 4.6]{dorfler2012exploring}, one can show (i) and (ii) by using the ultimate boundedness theorem \cite[Theorem 4.18]{khalil2002nonlinear} under condition \eqref{cluster_condition}.
\end{IEEEproof}

	Suppose there is only $1$ oscillator in each community (i.e., $N=1$), and it hold that $\mathcal T_r=\mathcal T_M$, $D_o=0$, Theorem \ref{theo:2-norm} reduces to the best-known result on the incremental $2$-norm in  single level networks {\cite[Theorem 4.6]{dorfler2012exploring}}.
	One observes that the established result in Theorem~\ref{theo:2-norm} is quite restrictive if the number of oscillators is large because we use the incremental $2$-norm metric. First, the critical value $\lambda_2(L)$ is quite conservative since the right side of \eqref{cluster_condition} depends greatly on the number of oscillator in the network. Second, the region of attraction we have identified in Theorem \ref{theo:2-norm}(ii) is quite small. To ensure $\| B_c^\top x(0) \|_2 <\gamma< \pi$, the initial phases are required to be nearly identical. In the next section, we use incremental-$\infty$ norm, aiming at obtaining less conservative results.

\section{Incremental $\infty$-Norm} \label{inft-norm}
\subsection{Main Results}

{In this subsection, we take the following function as a Lyapunov candidate for partial phase cohesiveness:}
\begin{align} \label{phas:infty}
	V(x)=\|B_c^\top x\|_\infty,
\end{align}
which is also referred to as the incremental $\infty$-norm metric. It evaluates the maximum of the pairwise phase differences, and thus does not depend on the number of oscillators. Then, one notices that $\mathcal S_\infty({\varphi})$ in \eqref{S_phi} can be rewritten into
\begin{align}\label{S:redefine}
	\mathcal S_\infty({\varphi})=\left\{\theta\in\mathbb{T}^{MN}:{V(x)=\|B_c^\top x\|_\infty} \le \varphi\right\}.
\end{align}	

 To the best of the authors' knowledge, the incremental $\infty$-norm has not been used to established explicit conditions for phase cohesiveness analysis in  weighted complete or non-complete networks, although some implicit conditions ensuring local stability of phase-locked solutions, such as \cite{dorfler2013synchronization,Jafarpour2018}, have been obtained. To obtain explicit conditions by using of the incremental $\infty$-norm, it is always required that the oscillators in a network have the same coupling structures (see~\cite[Theorem 6.6]{dorfler2014synchronization}, \cite{dorfler2011critical}). The oscillators in a non-complete network always have distinct coupling structures, which makes the analysis quite challenging.  {To overcome the challenge, we introduce the notion of the \textit{generalized complement} graph as follows, which can be viewed as an extension of the complement graph of an unweighted graph.}
\begin{definition}
	Consider the weighted undirected graph $\mathcal G$ with the weighted adjacency matrix $Z$, and let $K_m$ be the maximum coupling strength of its edges. Let $A_c$ denote the unweighted adjacency matrix of the complete graph with the same node set as $\mathcal G$.  We say $\bar{\mathcal G}$ is the \emph{generalized complement} graph of $\mathcal G$ if the following two are satisfied: 1) it has the same node set as $\mathcal G$; 2) the weighted adjacency matrix is given by $\bar Z:=K_mA_c-Z$.
\end{definition}

Let $K_m$ be the maximum element in the matrix \eqref{adjaceny_R}, and $A_c$ the unweighted adjacency matrix of the complete graph consisting of the same node set as $ \mathcal G_r$. Then $\bar Z=K_mA_c-Z$ is the weighted adjacency matrix of the generalized complement graph $ \bar{\mathcal G}_{r}$. In order to enable the analysis using the incremental $\infty$-norm, we then rewrite \eqref{dyna:xi} into the form taking the difference between the complete graph and the generalized complement graph
\begin{align*}
	\dot{x}_i=\varpi_i - K_m \sum_{n=1}^{Nr} &\sin (x_i-x_n) +\sum_{n=1}^{Nr} \bar z_{i,n}\sin(x_i-x_n) \nonumber\\
	&+\sum_{q=r+1}^{M}  \sum_{n=1}^{N} {a^{\mu(i),q}_{\rho(i),n}}\sin(\theta_n^q-x_i),
\end{align*}
where $i\in { \mathcal T_{Nr}}$. Accordingly, the incremental dynamics \eqref{incremental} can be rearranged into
\begin{align}\label{dyna:compl:increm}
	\dot{x}_i&-\dot{x}_j=\varpi_i -\varpi_j- K_m \sum_{n=1}^{Nr} \left(\sin (x_i-x_n)-\sin (x_j-x_n)\right)\nonumber\\ 
	&+\sum_{n=1}^{Nr} \left(\bar z_{in}\sin(x_i-x_n)- \bar z_{jn}\sin(x_j-x_n)\right)+u_{ij},
\end{align}
where $i,j\in {\mathcal T_{Nr}}$, and $u_{ij}$ is given by \eqref{incremental}. 

In the incremental $2$-norm analysis, the algebraic connectivity plays an important role since it relates to the matrix induced $2$-norm. When we proceed with the incremental $\infty$-norm  analysis, the corresponding ideas in terms of  the matrix induced $\infty$-norm are introduced subsequently.
Let $\bar D_m:=\|\bar Z\|_\infty$, and call it the \textit{maximum degree} of the generalized complement graph $\bar {\mathcal G}_r$. Let ${D_s^{\rm in}}:=\min_{i =1,\dots,Nr}\sum_{j=1}^{Nr}z_{ij}$, which we call the \textit{minimum internal} degree of $\mathcal G_r$. {Likewise, let the maximum external degree be $D_{m}^{\rm ex}:=\|D^{\rm ex}\|_\infty$.}
The following proposition provides a relation between the maximum degree of $\bar{\mathcal{G}}_r$ and minimum internal degree of $\mathcal G_r$. 
\begin{proposition}\label{propo:1}
	Given the graph $\mathcal G_r$, its minimum degree and the maximum degree of the associated generalized complement graph satisfy $\bar D_m=(Nr-1)K_m-D_s^{\rm in}$.
\end{proposition}
\begin{IEEEproof}
	From $\bar Z=K_m A_c-Z$, the following holds:
	\begin{align*}
		\bar{z}_{ij}=\left\{ \begin{array}{ll}
			0,&i=j\\
			K_m-z_{ij},&i\neq j.
		\end{array} \right.
	\end{align*}
	By taking the summation with respect to $j$, we have 
	\begin{align*}
		\sum\nolimits_{j =1} ^{Nr}\bar z_{ij}=(Nr-1)K_m - \sum\nolimits_{j =1}^{Nr}  z_{ij}, 
	\end{align*}
	where $z_{ii}=0$. 
	From the definition of the $\infty$-norm of the matrix and the fact that all the elements of $\bar Z$ and $Z$ are non-negative, it follows that 
	\begin{align*}
		\bar D_m=\|\bar Z\|_\infty & =\max_{i =1,\dots,Nr} \left( (Nr-1)K_m-\sum\nolimits_{j =1}^{Nr}  z_{ij} \right)\\
		&= (Nr-1)K_m-\min_{i =1,\dots,Nr} \sum\nolimits_{j =1}^{Nr}  z_{ij}\\
		&=(Nr-1)K_m-{D_s^{\rm in}}.
	\end{align*}
	The proof is complete.
\end{IEEEproof}

Now we provide our main result in this paper.

\begin{theorem}\label{Theo:inf-norm}
	{Suppose that the minimum internal degree ${D_s^{\rm in}}$ is greater than the critical value specified by
	\begin{align}
	{D_s^{\rm in}} > \frac{\|B_c^\top\varpi\|_\infty+2 {D_{m}^{\rm ex}}+ (Nr-2) K_m}{2}.\label{codi:Ds:1}
	\end{align}}
	Then, there exist two solutions, $\varphi_{s} \in [0,\pi/2)$ and $\varphi_{m} \in (\pi/2,\pi]$, to the equation $\|B_c^\top \varpi\|_\infty + 2 {D_{m}^{\rm ex}} + 2(Nr-1) K_m - 2{D_s^{\rm in}}=NrK_m\sin \varphi$, which are given by
	\begin{align}
		&\varphi_{s}=\arcsin \left( \frac{\|B_c^\top \varpi\|_\infty + 2 {D_{m}^{\rm ex}} + 2(Nr-1) K_m - 2{D_s^{\rm in}} }{NrK_m} \right), \label{phi:s}\\
		&\varphi_{m}=\pi-\varphi_{s},\label{phi:m}
	\end{align} 
	respectively. Furthermore, the following statements hold:
	\begin{enumerate}
		\item[(i)] For any $\varphi \in [\varphi_{s},\varphi_{m}]$, { $ {\mathcal S}_\infty (\varphi)$ is a positively invariant set of the system \eqref{model}};
		\item[(ii)] For every initial condition $\theta(0) \in \mathbb{T}^{MN}$ such that $\varphi_s<\|B_c^\top x(0)\|_\infty <\varphi_m$, the solution $\theta(t)$ to \eqref{model} converges to $ {\mathcal S}_\infty (\varphi_{s})$.
	\end{enumerate}	
\end{theorem}
\begin{IEEEproof}
	We first prove (i) by showing that  the {upper} Dini derivative of $V(x(t))$ along the solution to \eqref{model},
	\begin{align*}
		D^+V(x(t))
		= \mathop{\limsup}\limits_{\tau \to 0^+}\frac{V({x}(t+\tau))  - V({x}(t))} {\tau},
	\end{align*}
	satisfies $D^+V(x(t))\le 0$ when $V(x(t))=\varphi$. Define $\mathcal I'_M(t):=\{i: x_i(t)=\max_{j\in \mathcal{V}_r} x_j(t)\}$ and $\mathcal I'_S(t) := \{i: x_i(t)=\min_{j\in \mathcal{V}_r} x_j(t) \}$. Then one can rewrite \eqref{phas:infty} into
	\begin{align*}
		V(x(t) )=|x_p(t)-x_q(t)|, \ \forall p\in \mathcal I'_M(t), \forall q\in \mathcal I'_S(t).
	\end{align*}
	Let $\mathcal I_M(t):=\{i:\dot{x}_i(t)=\max_{j\in \mathcal I'_M}  \dot{x}_j(t)\}$ and ${\mathcal I_S}(t):=\{i:\dot{x}_i(t)=\min_{j\in \mathcal I'_S} \dot{x}_j(t)\}$. Then the upper Dini Derivative  is 
	\begin{align*}
		D^+V(x(t)) = \dot {x}_m(t)-\dot {x}_s(t)
	\end{align*}
	for all $m\in \mathcal I_M(t)$ and $s\in \mathcal I_S(t)$. It follows from  \eqref{dyna:compl:increm} that
	\begin{align*}
		&D^+V({x}(t))=\dot{x}_m - \dot{x}_s\\
		&=\varpi_m-\varpi_s-K_m\sum_{n=1}^{Nr}\left(\sin(x_m - x_n) -\sin(x_s-x_n)\right)\nonumber\\
		&+\sum_{n=1}^{Nr}\left(\bar z_{mn}\sin(x_m-x_n) -\bar z_{sn}\sin(x_s-x_n)\right) \nonumber +u_{ms}
	\end{align*}
	By using the trigonometric identity $\sin(x)-\sin(y)=2\sin \frac{{x-y}}{2} \cos \frac{x+y}{2}$,
	we have 
	\begin{align*}
		D&^+V({x}(t))=\varpi_m-\varpi_s\\
		&-2K_m\sum_{n=1}^{Nr} \sin\left( \frac{x_m-x_s}{2} \right) \cos \left( \frac{x_m - x_n}{2}- \frac{x_n - x_s}{2} \right) \nonumber\\
		&+\sum_{n=1}^{Nr}\left(\bar z_{mn}\sin(x_m-x_n) -\bar z_{sn}\sin(x_s-x_n)\right)+u_{ms} \nonumber .
	\end{align*}
	Since for any $\varphi \in [0,\pi]$, $V(x(t))=\varphi$ implies that $x_m(t)-x_s(t)=\varphi$, it follows that 
	\begin{align*}
		-\frac{\varphi}{2} \le \frac{x_m(t)-x_j(t)}{2} -\frac{x_j(t) - x_s(t)}{2} \le \frac{\varphi}{2}. 
	\end{align*}
	Consequently, {from the double-angle formula $\sin (\varphi)=2\sin(\varphi/2) \cos(\varphi/2)$}, it holds that 
	\begin{align*}
		D&^+V({x}(t)) \le \varpi_m-\varpi_s-NrK_m \sin (\varphi) \\
		&+\sum_{n=1}^{Nr}\left(\bar z_{mn}\sin(x_m-x_n) -\bar z_{sn}\sin(x_s-x_n)\right)+u_{ms}.
	\end{align*}
	Recalling the definitions of $\bar D_m$ and  ${D_{m}^{\rm ex}}$, one knows that 
	\begin{align*}
		\left|\sum\nolimits_{n=1}^{Nr}\left(\bar z_{mn}\sin(x_m-x_n) -\bar z_{sn}\sin(x_s-x_n)\right)\right|\le 2 \bar D_m
	\end{align*}
	and $|u_{ms}|\le 2{D_{m}^{\rm ex}}$. {As a consequence, from $\varpi_m-\varpi_s\le\|B_c^\top\varpi\|_\infty$ and Proposition \ref{propo:1}, we have
		\begin{align}
			D^+V({x}(t)) \le& \varpi_m-\varpi_s-NrK_m \sin (\varphi) +2\bar D_m+2 {D_{m}^{\rm ex}}\nonumber\\
			\le&  f(\varphi ),\label{Dini_inequ}
		\end{align}
		where
		\begin{align*}
			f(y ):=\|B_c^\top\varpi\|_\infty&- NrK_m \sin (y) \\
			&+ 2\big( (Nr-1)K_m-{D_s^{\rm in}}\big) + 2 {D_{m}^{\rm ex}}.
		\end{align*}
		
		Now, we aim to find a subinterval in $[0,\pi]$ such that $f(\varphi) \le 0$ for any $\varphi$ in it. If the  condition~\eqref{codi:Ds:1} holds, then $f(\pi/2) < 0$ and thus there exists such a subinterval around $\varphi=\pi/2$. Moreover, $\sin (y)$ is an increasing and decreasing function in $[0,\pi/2]$ and $[\pi/2,\pi]$, respectively. {Then there always exist two points $y_1 \in [0,\pi/2),y_2\in (\pi/2,\pi]$ such that $f(y_1)=f(y_2)=0$. These two points $y_1$ and $y_2$ are nothing but $\varphi_s$ in \eqref{phi:s} and $\varphi_m$ in \eqref{phi:m}, respectively.} In summary, for any $\varphi \in [\varphi_{s},\varphi_{m}]$, $D^+V(x(t))\le 0$ when $V(x(t))=\varphi$, which implies that  ${\mathcal S}_\infty (\varphi)$ is positively invariant.
				
	} 
			
	Next, we prove (ii). Given $x(0)$, it follows from \eqref{Dini_inequ} that for any $t$ there exists $\gamma(t)$ satisfying $\gamma(t)=V(x(t))$ such that 
	\begin{align}\label{D:ineq}
	D^+V(x(t))	\le &\|B_c^\top \varpi\|_\infty - NrK_m \sin (\gamma(t)) \nonumber\\
	&+ 2\left( (Nr-1)K_m-{D_s^{\rm in}} \right) + 2 {D_{m}^{\rm ex}}. 
	\end{align}
	{Recalling that the initial condition satisfies that $\varphi_s<\|B_c^\top x(0)\|_\infty <\varphi_m$, one knows that $\varphi_s < \gamma(0)<\varphi_m$. Then the right side of \eqref{D:ineq} is negative, and thus the strict inequality $D^+(V(x(0)))<0$ holds. From $t=0$ on, $D^+(V(x(0)))<0$ for all $t$ such that $\varphi_s<\gamma(t)<\varphi_{m}$, and $D^+(V(x(0)))\le 0$ if $\gamma(t)=\varphi_s$. One can then conclude that $\theta(t)$ converges to $\mathcal{S}_\infty (\varphi_s)$.}	
\end{IEEEproof}

	The following proposition provides a necessary condition for $K_m$ such that \eqref{codi:Ds:1} can be satisfied.
	\begin{proposition}\label{Condi:Km}
		 Suppose that ${D_s^{\rm in}}$ satisfies the condition \eqref{codi:Ds:1},  then $K_m$ satisfies the following inequality
		\begin{align}
		K_m>\frac{\|B_c^\top \varpi\|_\infty+2 {D_{m}^{\rm ex}}}{Nr}. \label{codi:Km:1}
		\end{align}
	\end{proposition}
	{\begin{IEEEproof}
		If the condition \eqref{codi:Ds:1} is satisfied, we have 
		\begin{align*}
		\|B_c^\top \varpi\|_\infty+2 {D_{m}^{\rm ex}} +(Nr-2)K_m <2{D_s^{\rm in}}.
		\end{align*}
		One notices that ${D_s^{\rm in}} \le (Nr-1)K_m$ since there are at most $Nr-1$ edges connecting each node, and the coupling strength of each of them is at most $K_m$. It then follows that 
		\begin{align*}
		\|B_c^\top \varpi\|_\infty+2 {D_{m}^{\rm ex}} +(Nr-2)K_m <2 (Nr-1)K_m,
		\end{align*}
		which implies $	K_m>\left({\|B_c^\top \varpi\|_\infty+2 {D_{m}^{\rm ex}}}\right)/{Nr}$.
	\end{IEEEproof}}
	
	In the study of synchronization or phase cohesiveness, the network is usually required to be connected. The following proposition shows that the condition~\eqref{codi:Ds:1} implies the connectedness of the graph  $\mathcal G_r$ since from the condition~\eqref{codi:Ds:1} the minimum internal degree satisfies ${D_s^{\rm in}}> (Nr-2)K_m/2$.

	\begin{proposition}\label{props:connected}
		Consider a graph $\mathcal G$ consisting of $n$ nodes. Let $K$ be the maximum coupling strength of its edges. Suppose the minimum degree of the nodes satisfies $D_s>{(n-2)K}/{2}$,
		and then the graph $\mathcal G$ is connected.
	\end{proposition}
	\begin{IEEEproof}
		We prove this proposition by contradiction. We assume that the graph is not connected, and let $i^*,j^*$ be any two nodes that belongs to two isolated connected components $\mathcal G_{i^*},\mathcal G_{j^*}$, respectively. Let the numbers of nodes that are connected to  $i^*,j^*$ be $n_{i^*}$ and $n_{j^*}$, respectively. The degree of $i^*$, denoted by $D_{i^*}$, satisfies
		\begin{align*}
		D_s\le D_{i^*} \le n_{i^*} K,
		\end{align*}
		It follows from the assumption $D_s>{(n-2)K}/{2}$ that $n_{i^*}>(n-2)/2$. which implies that the number of nodes in $\mathcal G_{i^*}$ is strictly greater than $n_{i^*}+1=n/2$. Likewise, one can show the number of nodes in $\mathcal G_{j^*}$ is strictly greater than $n_{j^*}+1=n/2$. Then the total number of nodes in these two isolated connected components is strictly greater $n_{i^*}+n_{j^*}+2=n$, which implies the number of node in the graph $\mathcal G$ is greater than $n$. This is a contradiction, and thus the network $\mathcal G$ is connected.	
	\end{IEEEproof}


\subsection{Comparisons}

We first compare the results in Theorems \ref{theo:2-norm} and \ref{Theo:inf-norm}. It is worth mentioning that the condition in Theorem \ref{Theo:inf-norm} is less dependent on the number of nodes $Nr$ than that in Theorem~\ref{theo:2-norm} in most cases. In sharp contrast to $\|B_c^\top \varpi\|_2$ and ${\| \hat B_c^\top D^{\rm ex}\|_2}$ in \eqref{cluster_condition},  both $\|B_c^\top \varpi\|_\infty$ and ${D_{m}^{\rm ex}}$ in \eqref{codi:Ds:1} are independent of $Nr$.  Specifically, if we take $\delta_s,\delta_m$ to be the smallest and largest elements in $|B_c^\top\varpi|$, respectively, it holds that $\delta_s\sqrt{Nr(Nr-1)/2} \le \|B_c^\top \varpi\|_2 \le \delta_m \sqrt{Nr(Nr-1)/2}$. A similar inequality holds for ${\| \hat B_c^\top D^{\rm ex}\|_2}$. Then, one can observe that $\|B_c^\top \varpi\|_2 + {\| \hat B_c^\top D^{\rm ex}\|_2}$ in \eqref{cluster_condition} can be much larger than $(Nr-2) K_m/2$ in \eqref{codi:Ds:1} if $Nr$ is large. More importantly, $\mathcal S_\infty(\varphi)$ is much larger than $\mathcal S_2(\varphi)$ for the same $\varphi$, which implies that the domain of attraction we estimated in Theorem~\ref{Theo:inf-norm} is much larger than that in Theorem~\ref{theo:2-norm}. Therefore, the convergence to a partially phase cohesive solution can be guaranteed by Theorem 2 even if the initial phases  are not nearly identical. 
	
	On the other hand, the condition \eqref{cluster_condition} can be less conservative than \eqref{codi:Ds:1}, but one would require the natural frequencies to be quite homogeneous, and meanwhile the external connections to be very weak in comparison with $K_m$. In addition, it can be observed  from Proposition \ref{props:connected} that each node in $ \mathcal G_r$ is required to have more than $(Nr-2)/{2}$ neighbors from the condition \eqref{codi:Ds:1}. In this sense, the condition \eqref{cluster_condition} is less conservative since it only requires $\mathcal G_r$ to be connected. 
	
	The following corollary provides a sufficient condition that is independent of the network scale for the partial phase cohesiveness in a dense non-complete subnetwork $\mathcal G_r$. It is certainly less conservative than its counterpart based on the  incremental $2$-norm.  
	\begin{corollary}
		Suppose each node in $\mathcal G_r$ is connected by at least $n_e$ edges, where $n_e>(Nr-2)/2$, and all the edges have the same weight $K$. Assume that
		\begin{align}
		K>\frac{\|B_c^\top\varpi\|_\infty+2 {D_{m}^{\rm ex}}}{2n_e-(Nr-2)}, \label{scale-free}
		\end{align}
		then the statements (i) and (ii) in Theorem \ref{Theo:inf-norm} hold. 
	\end{corollary}
	
	The proof follows straightforwardly by letting $D_s^{\rm in}=n_eK$ and $K_m=K$. Since $2n_e-(Nr-2)\ge 1$, any $K$ satisfying $K>\|B_c^\top\varpi\|_\infty+2 {D_{m}^{\rm ex}}$ satisfies the condition \eqref{scale-free} for any $Nr$.

Next, we compare our results with the previously-known works in the literature \cite{dorfler2011critical,dorfler2012exploring}. Since in the existing results, researchers usually deal with one-level networks, and study the complete phase cohesiveness, we assume, in what follows, that there is only one oscillator in each community in our two-level network, and let the set $ \mathcal T_r$ in which we want to synchronize the oscillators be the entire community set $\mathcal T_M$. Then we obtain the following two corollaries.

\begin{corollary}\label{Coro:2}
	Given an undirected graph $\mathcal G$, assume that the following condition is satisfied
	\begin{align}
	{D_s^{\rm in}} > \frac{\|B_c^\top  \varpi\|_\infty+(M-2)K_m}{2}, \label{Simpl:NComp}
	\end{align}
	then the solutions, $\varphi_{s} \in [0,\pi/2)$ and $\varphi_{m} \in (\pi/2,\pi]$, are respectively given by 
	\begin{align*}
	&\varphi_{s}=\arcsin \left( \frac{\|B_c^\top  \varpi\|_\infty + 2(M-1) K_m - 2 {D_s^{\rm in}} }{MK_m} \right), \\
	&\varphi_{m}=\pi-\varphi_{s}.
	\end{align*} 
	Furthermore, the following two statements hold:
	\begin{enumerate}
		\item[(i)] for any $\varphi \in [\varphi_{s},\varphi_{m}]$, the set ${\mathcal S}_\infty (\varphi)$ is positively invariant; 
		\item[(ii)] for every initial condition $x(0)$ such that $\varphi_s<|B_c^\top x(0)\|_\infty <\varphi_m$, the solution $\theta(t)$ converges to ${\mathcal S}_\infty (\varphi_{s})$ asymptotically.
	\end{enumerate}	
\end{corollary}
This corollary follows from Theorem \ref{Theo:inf-norm} by letting  $N=1$, $r=M$ and ${D_{m}^{\rm ex}}=0$. In this case, $K_m=\max_{i,j\in \mathcal T_M}a_{ij}$ is the maximum coupling strength in $\mathcal G$. Compared to the best-known result on the incremental $2$-norm \cite[Theorem 4.6]{dorfler2012exploring}, the result established in Corollary \ref{Coro:2} is {often}  less conservative. The explanation is similar to what we provide when we compare Theorem \ref{Theo:inf-norm} with Theorem \ref{theo:2-norm}. Assuming the network is complete, we obtain the following corollary. 

\begin{corollary}
	Suppose the graph $\mathcal G$ is complete, and the coupling strength is $K/M$. Assume that the coupling strength satisfies $K>{\|B_c^\top  \varpi\|_\infty}$. Then, $\varphi_{s}$ and $\varphi_{m}$ become
	\begin{align*}
	&\varphi_{s}=\arcsin \left( \frac{\|B_c^\top  \varpi\|_\infty  }{K} \right),
	&\varphi_{m}=\pi-\varphi_{s}.
	\end{align*}
	Furthermore, the statement (i) and (ii) in Corollary \ref{Coro:2} hold.
\end{corollary}

This result is actually identical to the well-known one found in \cite[Theorem 4.1]{dorfler2011critical}, which presents phase cohesiveness on complete graphs with arbitrary distributions of natural frequencies. 

\section{Numerical Examples}\label{simulation}
	In this section, we provide two examples to show the validity of the obtained results (see Example \ref{exam:partial}), and also to show their applicability to brain networks (see Example \ref{exam:brain}). We first introduce the order parameter as a measure of phase cohesiveness \cite{kuramoto1975self}, which is defined by $re^{i\psi}=\frac{1}{n} \sum_{i=1}^{n} e^{i \theta_j}$. The value of $r$ ranges from $0$ to $1$. The greater $r$ is, the higher the degree of phase cohesiveness becomes. If $r=1$, the phases are completely synchronized;  on the  other hand, if $r=0$, the phases are evenly spaced on the unit circle $\mathbb S^1$. 
	
	\begin{example}\label{exam:partial}
		We consider a small two-level network consisting of $6$ communities described in Fig.~\ref{Fig-1}. Each community consists of $5$ oscillators coupled by a complete graph. We assume that the oscillators between every two adjacent communities are interconnected in a way shown in Fig. \ref{Fig:inter}. The inter-community coupling strengths are given beside the edges in Fig.~\ref{Fig-1}. Denote $\omega=[{\omega^1}^\top,\dots,{\omega^6}^\top]^\top$, and let $\omega_1=0.5 ~{\rm rad/s}$ and $\omega_i=\omega_1+0.1(i-1)$ for all $i=2,\dots,30$. Let the local coupling strengths be $K^2=K^3=2.9$, and $K^1=K^4=K^5=K^6=0.01$. One can check that the condition \eqref{codi:Ds:1} is satisfied for the candidate regions of partial phase cohesiveness in the red rectangular, i.e., $\mathcal T_r=\{2,3\}$. The evolution of the incremental $\infty$-norm of the oscillators' phases in $\mathcal T_r$ is plotted in Fig. \ref{plot:Exa1:norm}, from which one can observe that starting from a value less than $\varphi_m$, $\|B_c^\top x(t)\|_\infty$ eventually converges to a value less than $\varphi_s$. One can then conclude that phase cohesiveness takes place between the communities $2,3$. On the other hand, it can be seen from Fig. \ref{plot:examp1:outside} that the value of $r$, which measure the level of synchrony, remains small, which means that the other communities in the network are always incoherent. These observations validate our obtained results on partial phase cohesiveness in Theorem \ref{Theo:inf-norm}.  Moreover, calculating the algebraic connectivity of the subgraph in the red rectangular, we obtain $\lambda_2(L)=5.6$, which is not sufficient to satisfy the condition \eqref{cluster_condition} in Theorem \ref{theo:2-norm}. Consistent with what we have claimed earlier, the results in Theorem \ref{Theo:inf-norm} can be sharper than those in Theorem \ref{theo:2-norm}.
			\begin{figure}[t!]
			\centering
			\subfigure[]{
				\includegraphics[scale=0.47]{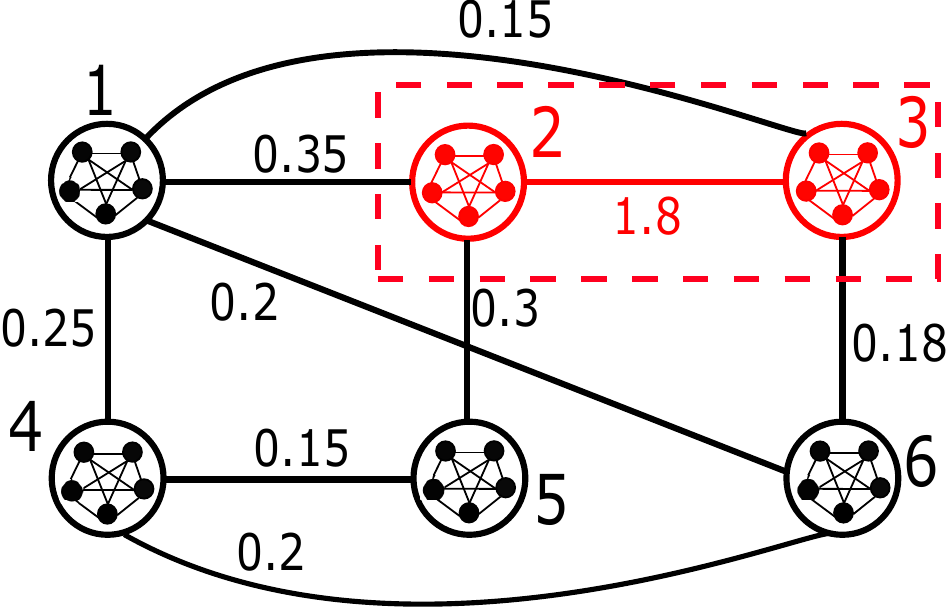}
				\label{Fig-1}}		
			\subfigure[]{\label{Fig:inter}
				\includegraphics[scale=0.22]{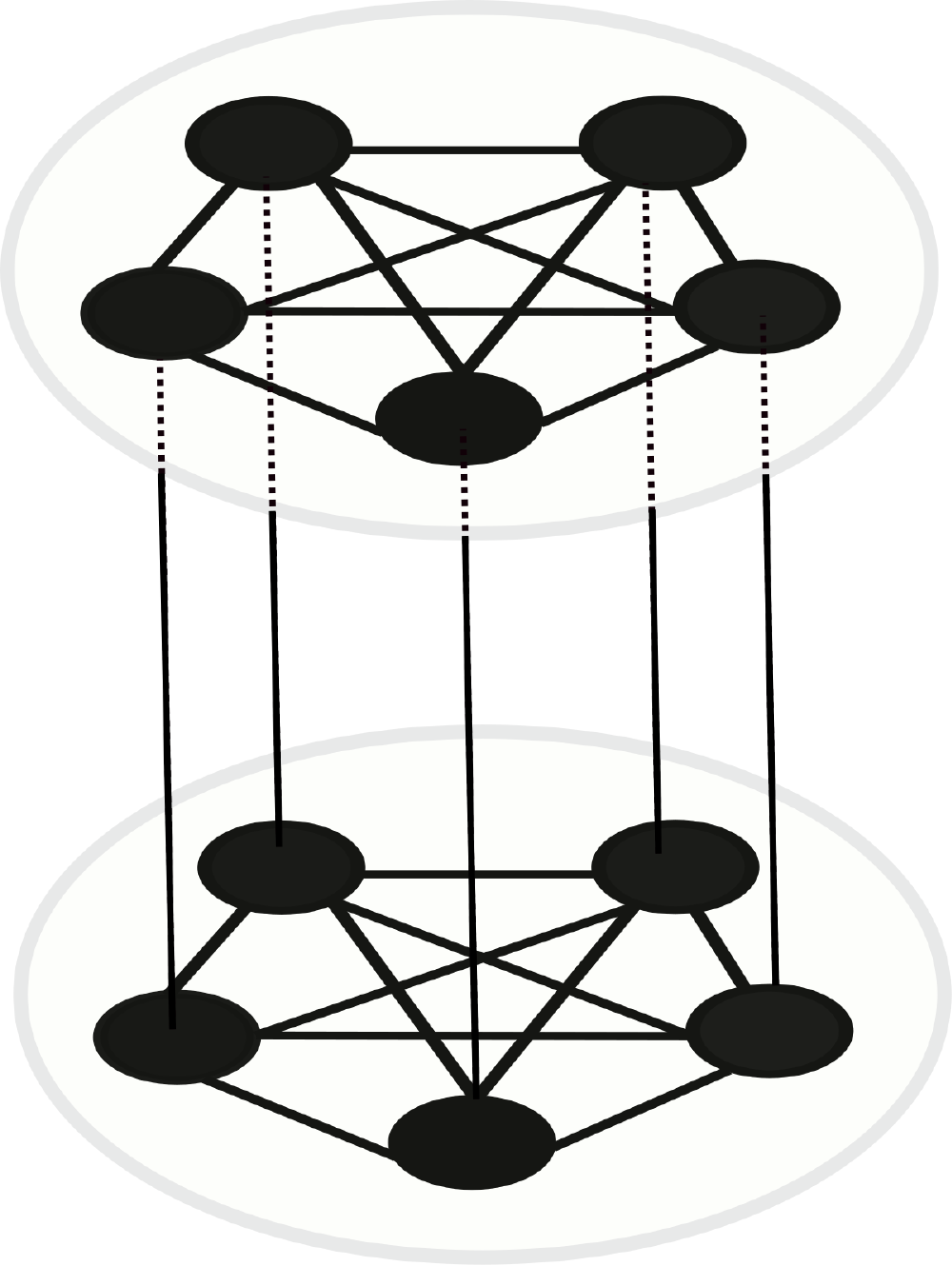}
				}		
			\subfigure[]{\includegraphics[scale=0.25]{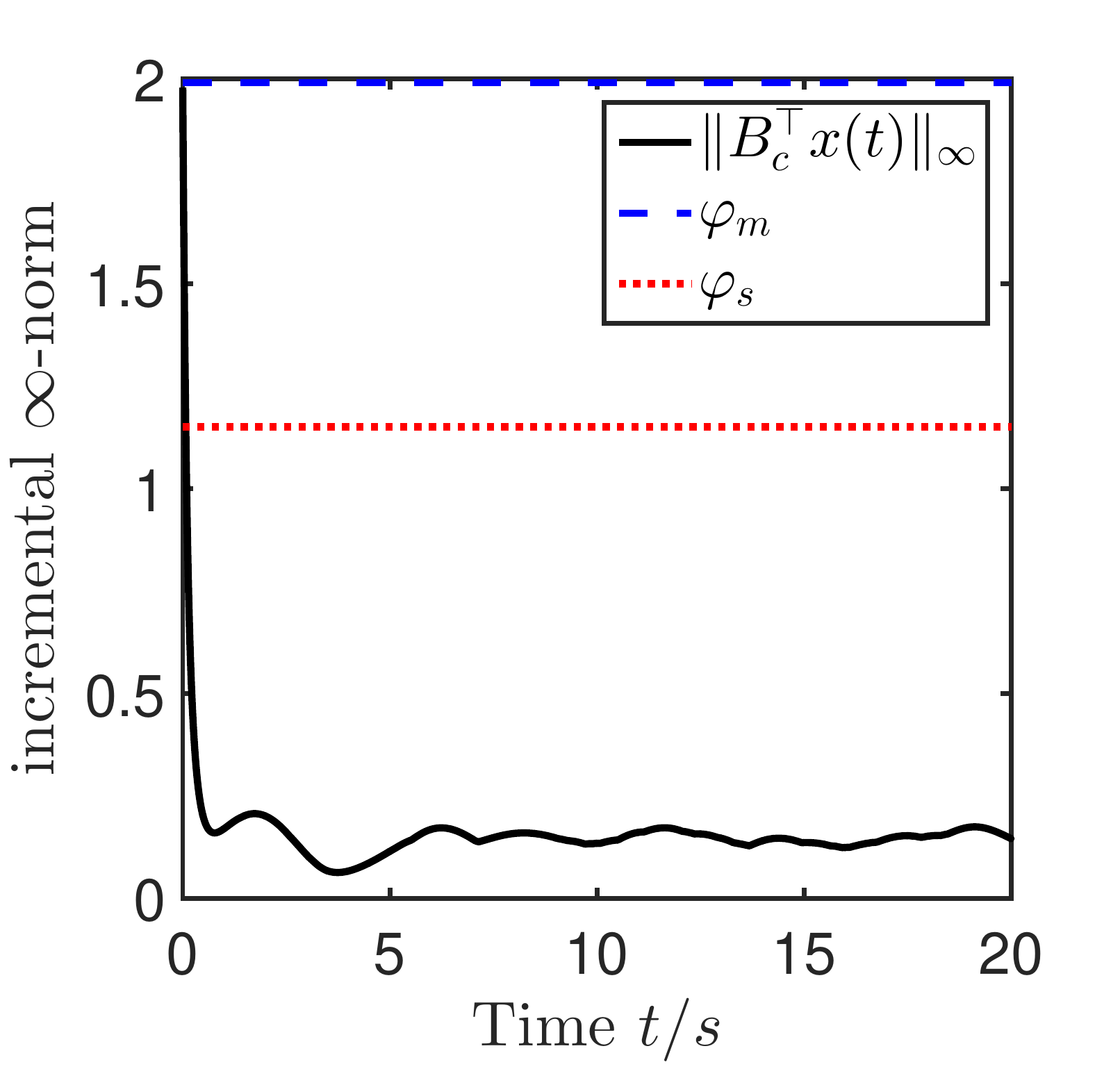}		
				\label{plot:Exa1:norm}			}
			\subfigure[]{\includegraphics[scale=0.25]{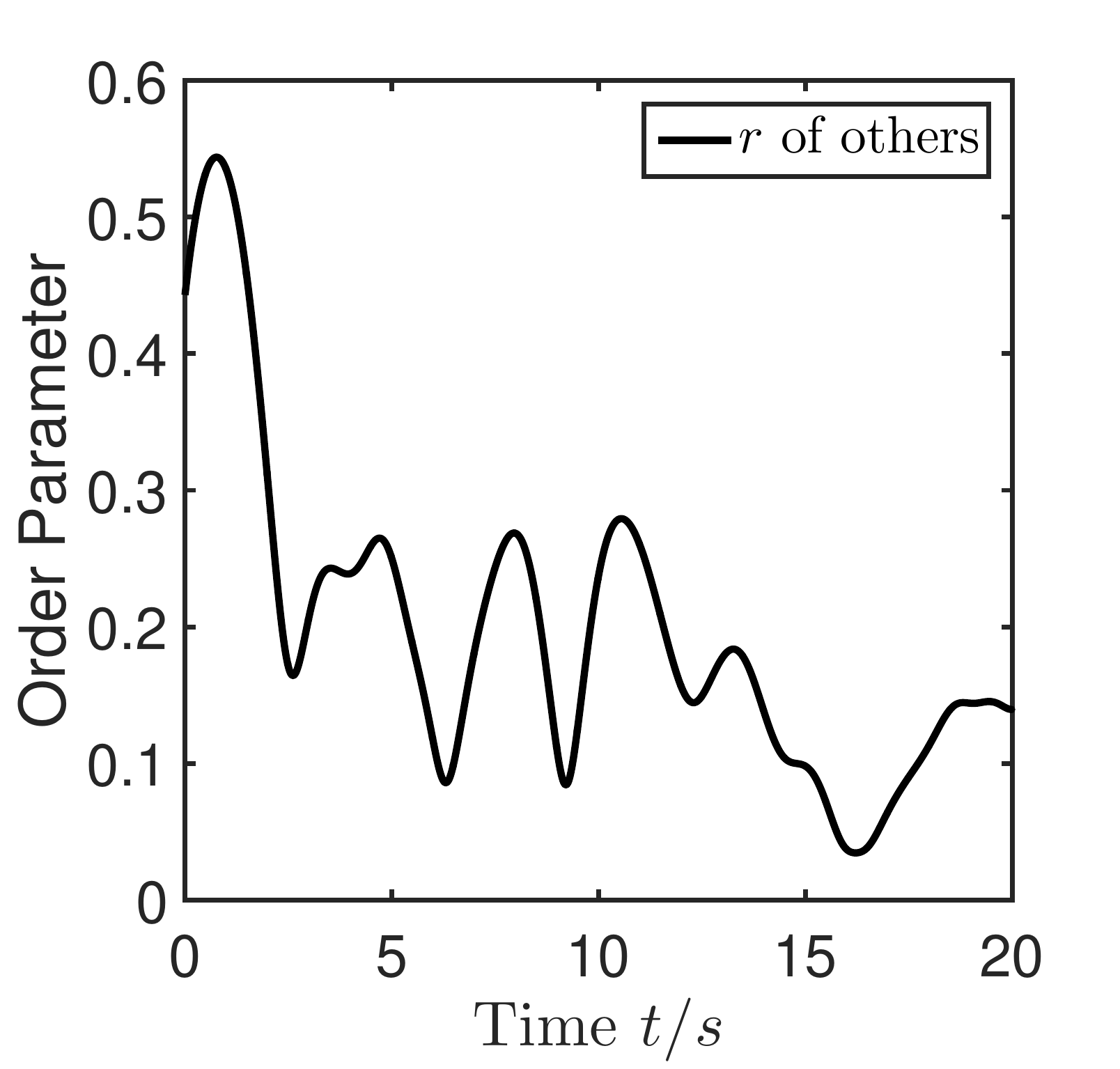}		
				\label{plot:examp1:outside}			}
			\caption{(a) The network structure  considered in Example \ref{exam:partial}; (b)  the interconnection structure: each oscillator in a community is connected to exact one oscillator in another; (c) the trajectory of $\|B_c^\top x(t)\|_\infty$, where $x=[\theta_j^p]_{10\times 1}, j=\mathcal T_{5}, p=2,3$; (d) the magnitude $r$ of the order parameter evaluated on other regions ($1, 4,5$ and $6$).}
		\end{figure}
	\end{example}


\begin{example}\label{exam:brain}
	\begin{figure}[t!]
		\centering
		\subfigure[]{
			\includegraphics[scale=0.42]{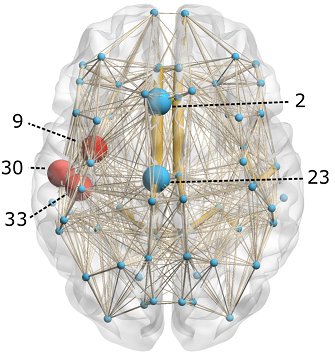}
			\label{Fig:brain}}
		\subfigure[]{
			\includegraphics[scale=0.26]{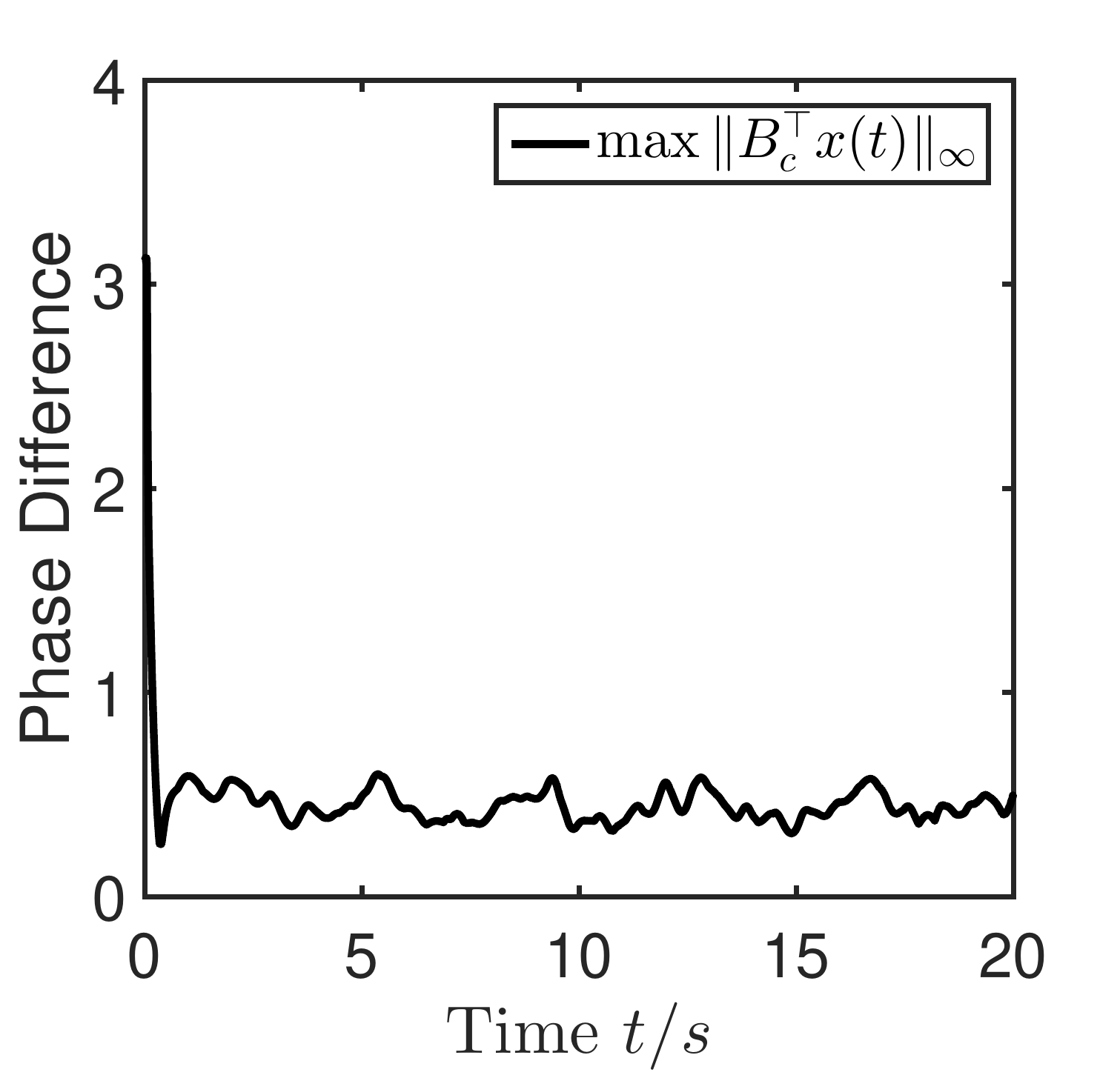}
			\label{Fig:phaseDBrain}}
		\subfigure[]{
			\includegraphics[scale=0.26]{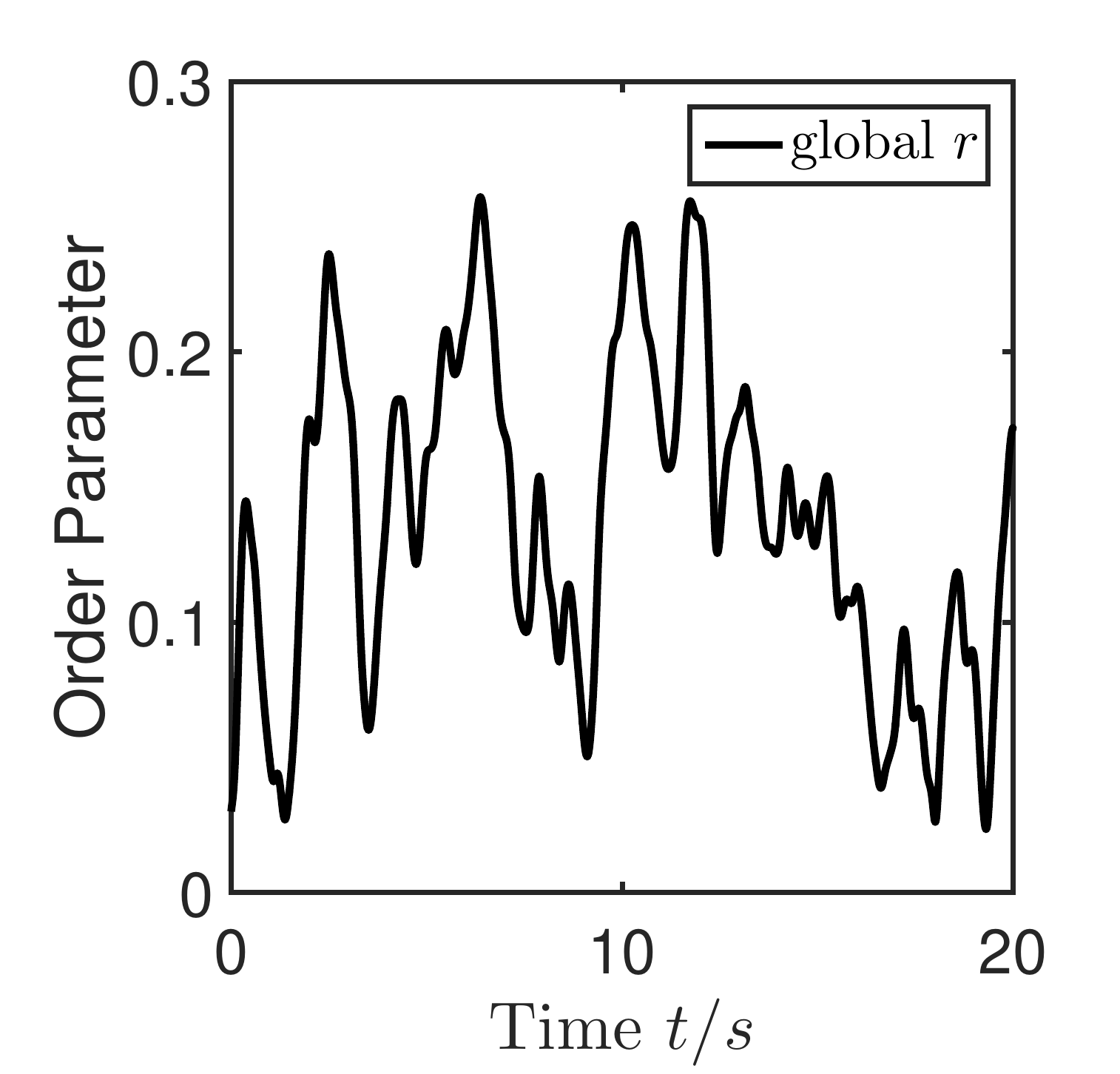}
			\label{Fig:OrderBrain}}
		\subfigure[]{
			\includegraphics[scale=0.26]{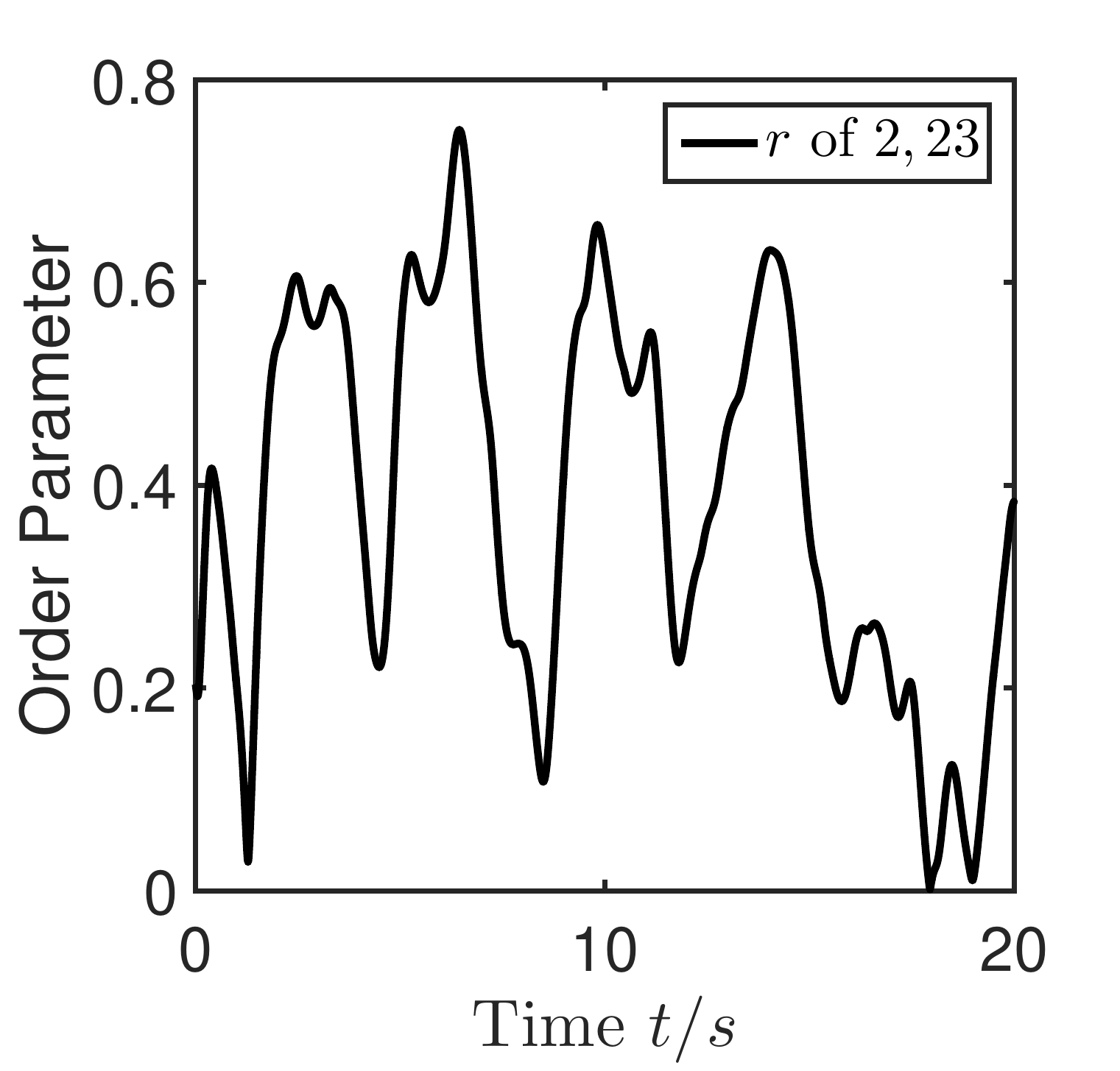}
			\label{Fig:2_23_OrPara}}
				
		\caption{(a) the anatomical brain network visualized by BrainNet Viewer \cite{xia2013brainnet}, edges only of weights larger than $0.15$ are shown for clarity;  (b) the maximum phase difference (absolute value) of the oscillators in $9,30,33$, where $x=[\theta_j^p]_{30\times 1}, j\in\mathcal T_{10}, p=9,30,33$; (c) the magnitude $r$ of the global order parameter; (d) the magnitude $r$ evaluated on the regions $2$ and $23$.} 
	\end{figure}
	In this example, we investigate partial phase cohesiveness in human brain with the help of an anatomical network consisting of $66$ cortical regions. The coupling strengths between regions are described by a weighted adjacency matrix $A=[a_{ij}]_{66\times 66}$ {whose elements represent axonal fiber densities computed by means of diffusion tensor imaging (DTI).} This matrix is the average of the normalized anatomical networks obtained from $17$ subjects \cite{finger2016modeling}. From our earlier analysis, strong regional connections play an essential role in forming partial phase cohesiveness. We identify some candidate regions by selecting the connections of strengths greater than $20$ ({visualized by the large size edges in Fig. \ref{Fig:brain}}). In particular, we consider two subsets of the brain regions $\{9,30,33\}$ and $\{2,23\}$, (see the red and blue nodes in Fig. \ref{Fig:brain}), and investigate whether phase cohesiveness can occur among them. 
	
	We use the model in which each of the $66$ regions consists of $10$ oscillators coupled by a complete graph with the coupling strength $K^p,p=1,\dots,66$, and any two adjacent regions are connected by $10$ randomly generated edges. {The weights of the $10$ edges connecting regions $i$ and $j$ are assigned randomly, and sum up to $a_{ij}$}. The natural frequencies of all the oscillators are drawn from a normal distribution with the mean $13\pi {\; \rm rad/s}$ ($6.5 {\;\rm Hz}$) and the standard deviation $1.5\pi$. Let the local coupling strengths $K^p=8$ for $p=9,30,33$, and $K^p=0.1$ for all the other $p$'s. Thus, we have obtained a two-level network from the anatomical brain network.  For this two-level network, we obtain some simulation results in Fig. \ref{Fig:phaseDBrain}, \ref{Fig:OrderBrain} and \ref{Fig:2_23_OrPara}. One can observe from Fig. \ref{Fig:phaseDBrain} that the regions $9,30,33$ eventually become phase cohesive, although the whole brain remains quite incoherent (see Fig. \ref{Fig:OrderBrain}, where the mean value of $r$ is approximately $0.15$). This observation indicates that strong regional connections can be the cause of partial phase cohesiveness.  On the other hand, one observes from Fig. \ref{Fig:2_23_OrPara} that without strong local coupling strengths phase cohesiveness does not take place between the regions $2$ and $23$ (the blue large nodes in Fig. \ref{Fig:brain}), although they have a strong inter-region connection, $a_{2,23}=52.8023$. This means that local coupling strengths could play an important role in selecting regions to be synchronized.
	
	From our theoretical results and simulations, we believe that there are at least two factors leading to partial brain synchronization. One factor relies on the anatomical properties of the brain network. The second factor depends on local changes of coupling strength. We hypothesize in this note that strong inter-regional coupling is one of the anatomical properties that allow for synchrony among brain regions. Then, selective synchronization of a subset of those strongly connected regions is achieved by increasing the local coupling strengths on the target regions, which can give rise to various synchrony patterns. Other properties of the anatomical brain network such as symmetries studied in \cite{nicosia2013remote} and \cite{QinCDC2018}, can be a topic of future work.
	\end{example}

\section{Concluding Remarks}

We have studied partial phase cohesiveness, instead of complete synchronization, of Kuramoto oscillators coupled by two-level networks in this note. Sufficient conditions in the forms of algebraic connectivity and nodal degree have been obtained by using the incremental $2$-norm and $\infty$-norm, respectively.  The notion of generalized complement graphs that we introduced provides a much better tool than those in the literature to estimate the region of attraction and ultimate level of phase cohesiveness when the network is weighted complete or uncomplete. However, the disadvantage of this method is that the number of edges connecting each node has a noticeable lower bound. The simulations we have performed provides some insight into understanding the partial synchrony observed in human brain. We are interested in investigating other mechanisms that could render partial synchronization. 

\ifCLASSOPTIONcaptionsoff
  \newpage
\fi

\bibliographystyle{IEEEtran}
\bibliography{IEEEabrv,references}

\end{document}